\documentclass[11pt,a4paper]{article}
\usepackage[letterpaper,includefoot,top=1in,bottom=1in,left=1in,right=1in]{geometry}
\usepackage{multirow}
\usepackage{float}
\usepackage{cite}
\usepackage[centertags,sumlimits]{amsmath}
\usepackage{amsfonts}
\usepackage{amssymb,graphics,psfrag}
\usepackage{array,epsfig,multirow,stmaryrd,graphicx}
\usepackage[all,cmtip]{xy}
\usepackage{mathrsfs}
\usepackage{color}
\usepackage{arydshln}  
\usepackage[bookmarks,breaklinks]{hyperref}
\usepackage{multirow}
\usepackage{eufrak}
\usepackage{verbatim}
\numberwithin{equation}{section}
\numberwithin{table}{section}
\usepackage{caption}
\usepackage{subcaption}
\usepackage{float}

\usepackage{bbm, dsfont}

\DeclareMathAlphabet{\mathbbold}{U}{bbold}{m}{n}

\definecolor{verde}{rgb}{0.05,0.6,0.33}
\definecolor{rojo}{rgb}{0.8,0.1,0}

\newcommand{\ArcCh}{\text{ArcCh}}

\newcommand{\beq}{\begin{equation}}
\newcommand{\eeq}{\end{equation}}
\newcommand{\be}{\begin{equation}}
\newcommand{\ee}{\end{equation}}
\newcommand{\bea}{\begin{eqnarray}}
\newcommand{\eea}{\end{eqnarray}}
\newcommand{\ben}{\begin{eqnarray*}}
\newcommand{\een}{\end{eqnarray*}}
\newcommand{\ba}{\begin{aligned}}
\newcommand{\ea}{\end{aligned}}
\newcommand{\bt}{\begin{tabular}}
\newcommand{\et}{\end{tabular}}
\newcommand{\bc}{\begin{center}}
\newcommand{\ec}{\end{center}}

\newcommand{\nn}{\nonumber}

\newcommand{\cref}{{\bf [check ref]}}

\newcommand{\tr}{\mathrm{Tr}\:}

\usepackage{amsmath}
\graphicspath{{./Pictures/}}

\entrymodifiers={+!!<0pt,\fontdimen22\textfont2>}

\begin{document}

\baselineskip=18pt
\setlength{\parskip}{6pt}

\begin{titlepage}

 \hspace{12cm} WITS-MITP-015
 \vspace{1.5cm}
 

\begin{center}


{\Large \bf Generalized entanglement entropy and holography}

\vskip 1cm


\normalsize {Nana Cabo Bizet}$^{a}$ \footnote{\texttt{nana.cabobizet@wits.ac.za}} and
\normalsize {Octavio Obreg\'on}$^{b}$ \footnote{\texttt{octavio@fisica.ugto.mx}},
  \vskip 0.0cm
    \emph{$^{a}$ Center for Theoretical Physics,  School of Physics, \\
 University of the Witwatersrand, 1 Jan Smuts Ave, Johannesburg, \\
  South Africa.} \\ [0.1cm]
 \emph{$^{b}$ Departamento de F\'{\i}sica, Divisi\'on de Ciencias e Ingenier\'{\i}as, Campus Le\'on, \\ [.1cm]
 Universidad de Guanajuato,Loma del Bosque 103, CP 37150,  Le\'on, Guanajuato, M\'exico}
\\

\end{center}
\vskip 0.0cm


\begin{center} {\bf ABSTRACT }
\end{center}

In this work, we first introduce a generalized von Neumann
entropy that depends only  on the density matrix.
This is based on a previous proposal by one of us  modifying the
Shannon entropy  by considering  non-equilibrium systems on stationary states, and
an entropy functional depending only on the  probability. We propose a generalization of
the replica trick  and find that the resulting modified von
Neumann entropy is precisely the previous mentioned entropy that
was obtained by other assumptions. Then, we address the question
whether alternative  entanglement entropies   can
play a role in the gauge/gravity duality. Our focus are 2d CFT and
their gravity duals. Our results show corrections to the
von Neumann entropy $S_0$ that  are  larger than the usual  $UV$ ones and also  than the corrections to the  length dependent $AdS_3$ entropy which
result comparable to the $UV$ ones. The  correction terms due to the new entropy  would modify the Ryu-Takayanagi  identification
between the CFT and the gravitational $AdS_3$  entropies.




\hfill \today
\end{titlepage}

\section{Introduction}

In previous works a generalized information entropy that depends
only on the probability distribution has been proposed
\cite{Obregon:2010zz,PhysRevE.88.062146}.  Considering non
equilibrium systems with a long-term stationary state that possess
a spatio-temporally fluctuating intensive quantity more general
statistics can be formulated called Superstatistics  \cite{super}.
These statistics first order corrections to the von Neumann entropy are the same for several relevant $f(\beta)$ distributions of the intensive quantity $\beta$ \cite{super,obregontorres}. By selecting the temperature as the fluctuating intensive quantity a  $\Gamma(\chi^2)$ distribution that depends on a parameter $p_l$
has been proposed.  From it, the corresponding Boltzmann factor
$B(E)$ was calculated and following \cite{tsallis2} the associated
entropy was constructed.  By introducing the appropriate
functional and by maximizing it the parameter $p_l$  can be
recognized as the probability distribution. This generalized
entropy can be expanded in series and has as a first term the well
known Shannon entropy \cite{Shannon}.  It is straightforward to
formulate the corresponding modified von-Neumann entropy
\cite{holzhey,calabresecardy,Calabrese:2009qy}.  It depends only
of the density matrix and its expansion has as a first term, the usual
von Neumann entropy.  As will be shown below, we propose a
generalization of the Replica trick by means of which this same
generalized von Neumann entropy arises.

We will focus on 2dCFT and  its gravity duals, in particular in
the proposal by Ryu-Takayanagi \cite{Ryu:2006bv,Ryu:2006ef}
which has been actively studied in recent years
\cite{Hubeny:2007xt,Nishioka:2009un,Casini:2011kv}. Works directed
to support the conjecture include
\cite{Fursaev:2006ih,Headrick:2010zt} and a calculation of it by a
gravitational entropy construction was given in
\cite{Lewkowycz:2013nqa} and for asymptotically $AdS_3$ pure
gravity in \cite{Faulkner:2013yia,Hartman:2013mia}. In addition
there has been a lot of recent progress in the study of entropies
for black holes \cite{Maldacena:2013xja} and higher derivatives
gravity \cite{Camps:2013zua}.


We will consider the ultraviolet corrections to the entanglement entropy of the 2d CFT \cite{holzhey}  and will calculate corrections to the $AdS_3$ length dependent entropy  \cite{Ryu:2006bv}.  The corrections on both sides are calculated necessarily from different assumptions and procedures.  They result however, in negative exponentials of the standard von Neumann
entropy $S_0$  multiplied by polynomials on $S_0$ and at each order the negative exponentials coincide for both entropies.  The generalized entanglement entropy \cite{Obregon:2010zz,PhysRevE.88.062146} we analyze in this work will depend  at each order in its expansion
of the whole von Neumann entropy with its corresponding $UV$ terms \cite{holzhey}.  When expanded, the most relevant terms are those that depend on the usual von Neumann entropy $S_0$
and the corrections also result in negative exponentials of it, multiplied by polynomials of the same $S_0$.  However, these exponentials result more relevant than those that arise in the $UV$ corrections
to the 2d CFT entanglement entropy and also to those arising in the expansion of the $AdS_3$ gravitational length dependent entropy.

In this work we will first in Section 2, briefly review  the well known entanglement entropies for a spatial subsystem $A$ in various 2 dimensional conformal field theories (2d CFT)with lattice spacing $a$ i.e. for a single interval, zero temperature and an infinite spatial dimension with and without boundary;  single interval, infinite spatial dimension and finite temperature; single interval, spatial cyclic dimension and zero temperature; multiple intervals with infinite spatial dimension and zero temperature\cite{holzhey,calabresecardy,Calabrese:2009qy}.  As known all of them have the same functional form.  In Section 3 we express the von Newmann entropy and propose  a
generalization of the Replica trick from which this modified entropy $S_+$ also arises.  For other possible generalized von Neumann entropies with corresponding generalized Shannon entropies \cite{Obregon:2010zz,PhysRevE.88.062146}  it is also possible to find the corresponding generalization of the Replica trick, we show the case that
also arises from a $\Gamma(\chi^2)$ distribution, basically by changing the sign of the parameter $p_l$.
In Section 4, the ultraviolet corrections to a 2d CFT entanglement entropy are reviewed and $S_{UV}$ results a function of $S_0$ and negative exponentials  of it multiplied by polynomials on $S_0$.  In Section 5 we calculate
the correction terms to the length dependent $AdS_3$  entropy $S_{grav}$.  We show that they are suppressed by negative exponentials of the same order as those in $S_{UV}$.  In Section 6 we pay attention to the generalized 2d CFT  entanglement
entropy $S_+$  \cite{holzhey,calabresecardy}, for the different theories, based on the proposed generalized Replica trick. They also depend on negative exponentials of  $S_0$    multiplied by polynomials of it.  The exponents of the exponentials are however smaller than those in $S_{UV}$ and
$S_{grav}$, being then larger and more relevant.

If these  more relevant corrections in this modified entropy should play a role, then, for example to maintain the Ryu-Takanayagi proposal one would need to define the appropriate corresponding entropy on the $AdS_3$ gravity. To find the correct
modified length entropy one would possibly need to define accordingly, modifications to the theory of gravitation.    Because $S_+$ has different origin and different
correction terms than $S_{UV}$ (and $S_{grav}$)  it would imply another generalized or corrected gravitation.  Nevertheless, a concrete formulation to find the appropiate modified area dependent $AdS_3$ entropy and its justification
in a modified  theory of gravitation is the matter of future research and is beyond the scope of this work.

\section{Review of entanglement entropies in 2d CFT}

Let us consider a relativistic, massless, lattice quantum theory in one space and one time dimensions,
we will review the cases considered in the work by Calabrese and Cardy  \cite{Calabrese:2009qy}.
The spatial coordinate $x$ is in the interval $x\in X$, $X=(0,L)$ or $X=[0,L)$ and the lattice spacing  is $a$.
\footnote{This gives a finite smearing at the end of the subsystems eliminating excitations
arbitrarily near to the boundary \cite{holzhey}.} The theory has
a Hamiltonian $\hat{H}$, and a complete set of observables $\{\phi(x)\}$. It constitutes
a conformal field theory in 1+1 dimensions. The density matrix for a thermal
state with inverse temperature $\beta$ is defined as $\rho(\{\phi''(x'')\}|\{\phi'(x')\})=Z(\beta)^{-1}\langle \{\phi''(x'') \}|e^{-\beta\hat{H}} | \{ \phi'(x') \}\rangle$
where $Z(\beta)=\tr e^{-\beta \hat{H}}$ is the partition function.

Let us define $A$ as a subsystem consisting of the points in the disjoint set of intervals
$A=(u_1,v_1)\cup....\cup(u_N,v_N)\in X$. A susbsystem $B$ is defined as
the complement of $A$. The reduced density matrix $\rho_A$ is computed
by tracing over the observables in $B$, i.e. $\rho_A=\tr_B \rho$.

Making n-copies of the system is possible to compute $\tr \rho_A^n$ for $n$
a positive integer, \cite{Calabrese:2009qy} what is the so called replica-trick. The partition function of  the system with this n-sheeted structure
reads $Z_n(A)$, with
\begin{equation}
\tr\rho^n_A=Z_n(A)/Z_1(A)^n.\label{replica1}
\end{equation}
The function is analytic in $n$ therefore admits a derivative which is also analytic
in $n$.  Computing the first derivative and making $n=1$ one obtains the von Neumann
entanglement entropy of the subsystem $A$ with the subsystem $B$. This magnitude is defined as
\begin{equation}
S_A=-\frac{\partial }{\partial n}\tr \rho_A^n|_{n=1}= - \tr \rho_A \log \rho_A.\label{22}
\end{equation}
Because of the analytic behaviour of $\tr \rho^n_A$
it is possible to derivate with respect to $n$ consecutive times obtaining always analytical
functions. This property will be used in the incoming sections.

For all the examples of 2d CFTs analyzed in [8] using the argument that the conformal transformation
property of certain two-point functions equals the conformal transformation of $\tr \rho_A^n$, for $a \rightarrow 0$,
one obtains \footnote{In this description the partition function normalization is such that it equals the partition function of the modular transformed system $\tau\rightarrow -1/\tau$ \cite{holzhey}.}
\begin{equation}
\tr \rho^n_A=c_n b^{(1/n-n)/6},\label{general}
\end{equation}
with $b$ a parameter depending on the model. In particular depends on the
spatial dimension $X$, the subsystem  $A$, the central charge $c$, the temperature $1/\beta$ and the boundary conditions. The results for the different cases can be read off in table \ref{modelos}.
There are various CFT in the ground state (zero temperature) and one CFT in a 
thermal mixed state (temperature $1/\beta$).

 \begin{table}[htdp]
\caption{ \label{modelos}Values of $\tr \rho_A^n$ in various CFTs with central charge $c$ \cite{Calabrese:2009qy}.}
\begin{center}
\begin{tabular}{|c|c|c|c|c|}\hline
case&1d system &Subsystem $A$ & Temperature& $\tr \rho_A^n$\\ \hline
I&$(0,\infty)$ &$(u,v)$ & $0$ & $c_n(l/a)^{-\frac{c}{6}(n-1/n)}$\\
II&$[0,\infty)$& $[0,l)$ & $0$ &$\tilde{c}_n(2l/a)^{-\frac{c}{12}(n-1/n)}$ \\
III&$(0,\infty)$& $(u,v)$ & $\beta^{-1}$ &  $c_n\left(\frac{\beta}{\pi a}\sinh \left(\frac{\pi l}{\beta}\right)\right)^{-\frac{c}{6}(n-1/n)}$\\
IV&$(0,L)$ periodic bc& $(u,v)$ & $0$ &  $c_n\left(\frac{\beta}{\pi a}\sin \left(\frac{\pi l}{L}\right)\right)^{-\frac{c}{6}(n-1/n)}$\\
V&$(0,\infty)$& $\cup (u_i,v_i)$ & 0 & $c_n^N \left(\frac{\prod_{j\leq k}(v_k-u_j)}{\prod_{j\leq k}(u_k-u_j)(v_k-v_j)}\right)^{-\frac{c}{6}(n-1/n)}$\\ \hline
\end{tabular}
\end{center}
\end{table}%

In appendix \ref{largeUV} we write the exact formula for one case I  $\tr \rho_A^n$  to show
that it corresponds to (\ref{general}) with $c_n=1$.

\section{Generalized entanglement entropies and replicas}

In this section we propose a generalization of the Replica trick to express the entanglement entropy $S_+$ \cite{Obregon:2010zz,PhysRevE.88.062146} for a subsystem $A$ as a series depending on all the n-replicas.


Expanding $S_+$ in powers one obtains an expression depending on $\tr \rho_A^n$ as
\footnote{The new statistic has a different dependence of the density matrix $\rho$ on the hamiltonian \cite{lenzimalacarne,PhysRevE.88.062146}. Calculations should then be performed taking these modifications to $\rho$
into account.  However, the corresponding correction terms to the entropy result smaller at  each order in the expansion.}
\begin{equation}
S_+=\tr (1-\rho_A^{\rho_A})=-\sum_{k\geq 1} \frac{1}{k!} \tr (\rho_A \ln \rho_A)^k,\label{S+}
\end{equation}
the modified Replica trick we propose is given by

\begin{equation}
S_{+}=-\sum_{k\geq 1} \frac{1}{k!}\lim_{n\rightarrow k} \frac{\partial^k}{\partial n^k}\tr \rho_A^n.\label{S+2}
\end{equation}
The last line can be computed in a general CFT with the use of the standard Replica trick as in (\ref{22}). It is possible to determine other generalized entanglement entropies by appropriately  further generalizing the Replica trick as in (\ref{S+2}); an example corresponds to the same $\Gamma(\chi^2)$ distribution considered in \cite{Obregon:2010zz,PhysRevE.88.062146} and is obtained by changing the sign of probabilities. The appropriate generalized
Replica trick would then be
\begin{eqnarray}
S_{-}=\tr (\rho_A^{-\rho_A}-1)=-\sum_{k\geq 1} \frac{1}{k!} (-1)^k\tr (\rho_A \ln \rho_A)^k,\\
=-\sum_{k\geq 1}(-1)^k \frac{1}{k!}\lim_{n\rightarrow k} \frac{\partial^k}{\partial n^k}\tr \rho_A^n . \nn
\end{eqnarray}

In Section 6 we will focus on the generalized entanglement entropy $S_+$.    In the next two sections we will
review the ultraviolet corrections to the 2d CFT entanglement entropy and find the corrections to the $AdS_3$ length dependent entropy proposed in \cite{Ryu:2006bv}.

\section{UV corrections to the 2dCFT entanglement entropy}

In this section we review the ultraviolet correction to the 2d CFT entanglement entropy.  We consider the specific CFT of free massless bosons in a system with spatial dimensions of length $L \rightarrow \infty$ (case I in table \ref{modelos}).  The subsystem is given by  the region $A=(0,l)$ and we consider a $UV$ cutoff $a$,
corresponding to the lattice spacing.


It is possible to make two consecutive conformal transformations that map the region of $z=\sigma+i \tau$
with $z\in (0,L)$ and $\tau\in (0,\infty)$ to a torus and then replicate the system to compute the partition
function $Z_n(A)$ \cite{holzhey}. The modular parameter of the torus is given by $\tau_1=i\pi x$, with $x=\ln (l/a)$.

On the replicated system  $\tau_n=i\pi n x$ is the modular parameter, and a modular transformation is performed
to obtain the modular transformed $\hat{\tau}_n=-1/\tau_n$
parameter with
\begin{eqnarray}
\hat{q}_n=e^{2\pi i\hat{\tau_n}},\,\hat{\tau}_n=i/(\pi n x),\, \frac{\partial}{\partial n}\hat{\tau_n}=-\frac{\hat{\tau_1}}{n^2}.
\end{eqnarray}
Computing the partition function
of the system of replicas on the modular transformed scheme $\hat{\tau}_n$ we obtain \footnote{We have employed only the leading contributions to the partition function
on the torus \cite{holzhey}.}
\begin{eqnarray}
\tr \rho_A^n=\frac{Z_n(A)}{Z_1(A)^n}&=&\frac{\eta(\hat\tau_1)^n\bar{\eta}(\hat{\bar{\tau_1}})^n}{\eta(\hat\tau_n)\bar{\eta}(\hat{\bar{\tau_n}})}.
\end{eqnarray}
In this scheme the limit $a\rightarrow 0$ corresponds to the limit $q=\hat q_1\rightarrow 0$.

 Using the expression for the derivative of the $\eta$ function $\frac{d \eta(\tau)}{d \tau}=\frac{\pi i}{12}\eta(\tau)E_2(\tau)$  one gets
\begin{eqnarray}
\frac{\partial \tr \rho^n}{\partial n}|_{n=1}&=&\left(\ln \eta_1(\hat{\tau})\bar{\eta_1}(\hat{\bar{\tau}})-\frac{\pi i}{12} E_2(\hat{\tau})/\tau+\frac{\pi i}{12} \bar{E}_2(\hat{\bar{\tau}})/\bar{\tau}\right),\label{eq12} \\
&=&\ln (q\bar{q})^{\frac{1}{24}}\prod_{n=1}^{\infty}(1-q^n)(1-\bar{q}^n)\nn \\
&-&\frac{\pi i}{12\tau}\left(1+\frac{2}{\zeta(-1)}\sum_{n}\sigma_1(n)q^{n}\right)
+\frac{\pi i}{12\bar{\tau}}\left(1+\frac{2}{\zeta(-1)}\sum_{n}\bar{q}^{n}\sigma_1(n)\right),\nn\\
&=&\ln (q)^{\frac{1}{6}}+\sum_{n\geq 1}  2\ln(1-q^n)
-2\sum_{n\geq 1}\frac{nq^{n}\ln q}{1-q^n},\nn\\
&=&-S_{UV} .\nn
\end{eqnarray}
The first term corresponds to the standard von Neumann entropy for this case $S_0= - \frac{1}{6} \ln q$.
The expansion of (\ref{eq12}) can be written as \cite{holzhey}
\begin{eqnarray}
S_{UV}&=&-1/6 \ln q-2 \left(1+\ln \frac{\partial}{\partial \ln q} \right)\sum_{k=1}^{\infty} \ln (1-q^k),\label{Suv}\\
 &=&S_0+2\sum_{k=1}^\infty\sum_{n=1}^\infty \left(\frac{1}{n}-6 S_0 k\right) e^{-6 S_0 k n},\nn\\
 &=&S_0 + 2  e^ {-6 S_0}  (1-6 S_0) + 2 e^{-12S_0} (3/2-18S_0) +  ...\nn
 \end{eqnarray}


\section{Corrections to the $AdS_3$ entropy}

Now we want to look at the entanglement entropy in $AdS_3$ \cite{Ryu:2006ef} which is proportional to the area of a minimal surface. In this case the length of an static geodesics determines its value. The $AdS_3$ metric depends on global coordinates $(t,\rho,\theta)$ as $ds^2=R^2(-\cosh \rho^2 dt^2+d\rho^2+\sinh\rho^2d\theta^2)$ \cite{Ryu:2006bv}. $R$ is the radius of $AdS_3$. A geodesic going from
radius $\rho_0$ and angle $\theta=0$ to radius $\rho_0$ and angle
$\theta=\frac{\pi l}{L}$ has a length given by  \cite{Ryu:2006ef}
\begin{equation}
L_{\gamma_A}=R*\ArcCh \left(1-Sin^2\left(\frac{l \pi}{L}\right)+\frac{1}{2}e^{2\rho_0}Sin^2\left(\frac{l \pi}{L}\right)
+\frac{1}{2}e^{-2\rho_0}Sin^2\left(\frac{l \pi}{L}\right) \right).
\end{equation}
We make use of the following expansion around the cutoff $\rho_0\rightarrow \infty$ with parameter $r=e^{2\rho_0}$:
\begin{eqnarray}
\ArcCh[1 - A + \frac{r A}{2} + \frac{A}{2 r}]&=&Log[r A] + \frac{(2-2 A)}{A r} - \frac{( A-3) (A-1)}{A^2 r^2} \\
&-& \frac{2 ((A-1) (10 + (A-8 ) A))}{3 A^3 r^3}+... ~ ~ .\nn
\end{eqnarray}
This gives us as a result {\it UV} corrections to the CFT entanglement entropy computed holographically by Ryu-Takayanagi \cite{Ryu:2006bv} on the gravity side
\begin{eqnarray}
S_{grav}&=&S_0+ \frac{c}{6e^{6 S_0/c}} \left(2 - 2 Sin^2\left(\frac{l\pi}{L}\right)  \right)\label{Sgrav}\\
 &-& \frac{c}{6e^{12 S_0/c}} \left(Sin^2\left(\frac{l \pi}{L}\right)-3\right) \left(Sin^2\left(\frac{l \pi}{L}\right)-1\right) \nn \\
&-& \frac{c}{6e^{18 S_0/c}} \left(Sin^2\left(\frac{l \pi}{L}\right)-1\right) \left(10+(Sin^2\left(\frac{l \pi}{L}\right)-8)Sin^2\left(\frac{l \pi}{L}\right)\right)+... ~ ~ ,\nn
\end{eqnarray}
with
\begin{equation}
S_0 = \left(\frac{R}{4G} \ln e^{2 \rho_0} Sin^2\left(\frac{l \pi}{L}\right)\right) . \nn
\end{equation}

The cutoff on  both sides is mapped via

\begin{equation}
e^{\rho_0}\equiv L/ \pi a,\, \frac{R}{4G}\equiv \frac{c}{6} , \label{map}
\end{equation}
where $a$, $L$ and $c$ are the lattice spacing, length and central charge of the 2d CFT
respectively. The further terms in this $S_{grav}$,  are also exponentially suppressed as is the case with the terms due to the {\it UV} contribution to the entanglement entropy of the CFT  (\ref{Suv}).


For an infinite spatial dimension and $c=1$ $S_{grav}$ can be approximated and it results in

\begin{equation}
S_{grav} \simeq  S_0 + \frac{1}{3} e^{-6 S_0} - \frac{1}{2}e^{-12S_0}  + ... ~~ .\label{Sgaprox}
\end{equation}

As stated in \cite{Ryu:2006bv}  the first term in the 2d CFT entanglement entropy (\ref{Suv}), under the identification (\ref{map}) coincides with the one in (\ref{Sgaprox}).
The two calculations are, however based in different procedures and assumptions.  Nevertheless, their corresponding correction terms result in expansions
in the same dominant negative exponentials, of the von Neumann entanglement entropy.  We will now proceed in the next section 6 to calculate the correction
terms to the entropy $S_+$ we propose (\ref{S+}) by means of our generalized Replica trick. We will show that even though they depend also on negative powers of
$S_0$ exponentials, these are smaller and consequently the correction terms should be more relevant.

\section{Generalized entanglement entropy $S_+$ and its $S_0$ dependence}
Here we give the expressions for the leading  terms, at each order of the $S_+$
expansion for a 2d CFT with $\tr \rho_A^n$ given by (\ref{modelos}) and $(\ref{general})$
considering a constant $c_n$ with negligible $\partial_n$ derivatives.

The first derivative (\ref{eq12}) of $\tr \rho^n$ evaluated in $n=1$
\begin{eqnarray}\label{eq1}
\frac{\partial }{\partial n}\tr \rho^n |_{n=1}=-\frac{1}{3} \ln b ,
\end{eqnarray}
gives minus the von-Neumann entropy $-S_0$. In order to determine $S_{+}$, we use the proposed generalized Replica trick (\ref{S+2}),
to get the following derivatives
\begin{eqnarray}\label{eq2}
\frac{\partial^2 \tr \rho^n}{\partial n^2}|_{n=2}=-\frac{e^{-3S_0/4}}{8 }S_0(1 + \frac{25}{8} S_0),
\end{eqnarray}
and
\begin{eqnarray}
\frac{\partial^3 \tr \rho^n}{\partial n^3}|_{n=3}=S_0 e^{-4 S_0/3}\left(\frac{1}{27}+\frac{5}{181} S_0+\frac{125}{729} S_0^2\right).\label{eq3}
\end{eqnarray}
From which we observe that the generalized entropy $S_+$, which includes the negative sum of (\ref{eq1}), (\ref{eq2}) and (\ref{eq3}) plus further terms,  has as dominant contribution $S_0$ and the corrections are exponentially suppressed polynomials on $S_0$.

As already mentioned in the Introduction, each term of the entropy $S_+$ depends in general also on all of  the $UV$ correction terms in $S_{UV}$.
One should calculate accordingly the terms in (\ref{S+2}).  For case I in table \ref{modelos} the first one is given by  (\ref{eq12}) and (\ref{Suv}). The second one is shown below
\begin{align}
\frac{\partial^2 \tr \rho^n}{\partial  n^2}|_{n=2}=&\frac{(\eta_1\bar{\eta_1})^2}{(\eta_2\bar{\eta_2})}\left((-S_{UV}+\frac{\pi i}{12}(E_2(\hat{\tau}_2)/n^2-E_2(\hat{\tau}_1))\hat{\tau}_1+c.c)^2 \right.\label{eq22} \\
 + &\frac{\pi i}{12} E_2(\hat{\tau}_2)\left(\frac{-2\hat{\tau}_1}{n^3}\right)-\left.\frac{\pi i}{12} E_2'(\hat{\tau}_2)\left(\frac{\hat{\tau}_1}{n^2}\right)^2+c.c. \right)|_{n=2},\nn\\
=&\hat q^{1/8}\prod_n(1+\hat q^{n/2})^4(1-\hat q^n)^2\nn \\
\times&\left(\left(-S_{UV}+\frac{\ln \hat q}{12}\left(-3/4+24\sum_{k,l\geq1}k  \hat q^{kl/2}(\hat q^{kl/2}-1/4)\right)\right)^2\right.\nn \\
-& \left. 2\left( \frac{1}{4}\sum_{k,l\geq 1} k^2 l \hat \hat q^{kl}\ln \hat q ^2-\frac{\ln q}{8\times 6}(1-24\sum_{k,l\geq 1}k \hat q^{kl})\right)\right),\nn
\end{align}
and the next two orders can be found in appendix \ref{eisenstein}.

We note that the dominant corrections due to the generalized entropy (\ref{eq2}) and (\ref{eq3}) are always negative exponentials of $S_0$ that multiply polynomials of it and they can be compared with the usual $UV$
corrections to the entanglement entropy $S_0$ (\ref{Suv}) and at the same time with the correction terms to the length dependent $AdS_3$ entropy (\ref{Sgaprox}).  The leading exponential terms of the $UV$
corrected entanglement entropy (\ref{Suv}) and the $AdS_3$  length dependent entropy (\ref{Sgaprox}) diminish with the exponent $-6n S_0$.  On the other hand, the first four leading correction terms in $S_+$
 diminish with exponents correspondingly  in the exponentials as $-\frac{3}{4}, -\frac{4}{3}, -\frac{15}{8} - \frac{12}{5}$, this can be checked in (\ref{eq2}), (\ref{eq3}) and in the appendix \ref{eisenstein} formulae. These terms are then more relevant than those
in (\ref{Suv}) and (\ref{Sgaprox}) and have another origin based on other assumptions   \cite{Obregon:2010zz,PhysRevE.88.062146}  and (\ref{S+2}).


\section{Further remarks}

The correction $UV$ terms to the entanglement entropy of the 2d
CFT (\ref{Suv}) arise with negative exponentials of the same
order than the correction terms to $S_{grav}$ (\ref{Sgaprox}).
The Ryu-Takayanagi  proposal\cite{Ryu:2006bv}, by means of the
mapping (\ref{map}), clearly identifies the first term in each
expansion resulting to be the same.  However, for the remaining
terms the leading negative exponentials result to be of the same
order in the exponentials for these two entropies.    As shown in
Section 6, the correction exponential terms in the generalized
entanglement entropy $S_+$, proposed by means of other assumptions
\cite{Obregon:2010zz,PhysRevE.88.062146}, (\ref{S+2}), are more
relevant and are of different order in the exponentials.  If in
these terms we would also interpret $S_0$ as the length in $AdS_3$;
the resulting entropy is no more a consequence of the corrections
to the $AdS_3$ area prescription.  The expected results will be
given by (\ref{eq2}) and (\ref{eq3}) (and further terms)  with
$S_0$ as the  area.  One would associate the usual corrections on
both sides $S_{UV}$ (\ref{Suv})   (and $S_{grav}$  (\ref{Sgrav}))
to indicate appropriate corresponding modifications to general
relativity.  The entanglement entropy analyzed  and
proposed in this work with its corrections to the von Neumann
entropy would imply another kind of modifications to gravity and a
modified $AdS_3$ length dependent entropy. Based on the
gravitational entropy \cite{Lewkowycz:2013nqa} and in the studies
for higher derivative gravity \cite{Camps:2013zua} it would be
interesting to explore what kind of gravitation arises in an
holographic context if the field theory is governed by a modified
entropy functional.  The generalized entropy $S_+$ as well as
other non-linear and non-equilibrium entropies could lead via
holography to new modified theories of gravity. This study is
beyond the scope of this work and is the matter of future
research.

\subsection*{Acknowledgements}

We would like to thank Alejandro Cabo Bizet, Jos\'e Luis L\'opez, 
Suresh Nampuri and \'Alvaro V\'eliz Osorio for useful discussions.
N. Cabo thanks the support of the NRF of South Africa,  
PROMEP, CEADEN and ``PNCB: Teor\'{\i}a Cu\'antica de Campos en F\'{\i}sica de Part\'{\i}culas y de la
Materia Condensada'' ICIMAF, CITMA. O. Obreg\'on thanks the
support by CONACYT project 135026, PROMEP and UG projects.

\newpage

\appendix

\section{Large UV cutoff $\tr \rho^n$}
\label{largeUV}

In the 2d CFT case I in table (\ref{modelos}) the value of $\tr\rho^n$ is given by
\begin{eqnarray}
\tr \rho^n&=&\frac{(\eta(\tau_1)\bar{\eta}(\bar{\tau}_1))^n}{\eta(\tau_n)\bar{\eta}(\bar{\tau}_n)}=\frac{n \ln q^{n-1}}{(2\pi)^{n-1}}(q\bar{q})^{(n-1/n)/24}\prod_k\frac{(1-q^k)^n(1-\bar{q}^k)^n}{(1-q^{k/n})(1-\bar{q}^{k/n})}.
\end{eqnarray}

In the limit of  $q\rightarrow 0$, which correspond to small lattice spacing we obtain
\begin{eqnarray}
\lim_{UV} \tr \rho^n=\frac{n \ln q^{n-1}}{(2\pi)^{n-1}}(q)^{(n-1/n)/12},\label{A2}
\end{eqnarray}
expression that includes terms which are absent in the result of the Calabrese and Cardy work  \cite{Calabrese:2009qy}. If the normalization neglects the sub-leading terms
on (\ref{A2}), which is equivalent to consider the torus partition function in the modular transformed system
as  $Z=1/(\eta\bar\eta)$ we obtain
\begin{eqnarray}
\tr \rho^n&=&\frac{(\eta(-1/\tau_1)\bar{\eta}(-1/\bar{\tau}_1))^n}{\eta(-1/\tau_n)\bar{\eta}(-1/\bar{\tau}_n)}=(q\bar{q})^{(n-1/n)/24}\prod_k\frac{(1-q^k)^n(1-\bar{q}^k)^n}{(1-q^{k/n})(1-\bar{q}^{k/n})}.
\end{eqnarray}
In the particular CFT of interest we have
\begin{eqnarray}
\tr \rho ^n&=&c_n (l/a)^{-c/6(n-1/n)},\nn
\end{eqnarray}
with $c_n=1$.


\section{Eisenstein series derivatives}

\label{eisenstein}


Drivatives of the $\eta-$function and the Eisenstein series $E_2$ are given by
\begin{align}
\frac{d \eta(\tau)}{d \tau}&=\frac{\pi i}{12}\eta(\tau)E_2(\tau),\, \\
\frac{d E_2}{d \tau}&=\frac{\pi i}{6}(E_2^2-E_4),\, \frac{d E_4}{d \tau}=\frac{2\pi i}{3}(E_2E_4-E_6),\, \frac{d E_6}{d \tau}=\pi i(E_2E_6-E_4^2),\nn \\
E_2''&=-\frac{\pi^2}{3\times 6}(E_2^2-3 E_2 E_4+2E_6),\nn\\
E_2'''&=-\frac{\pi^3 i}{3\times 6}\left(\frac{1}{3}(E_2^3-E_2E_4)-\frac{5}{2}E_2^2E_4+4E_2E_6-\frac{3}{2}E_4^2\right).\nn
\end{align}

Let us also gather explicit values of the series that we use
\begin{eqnarray}
E_2(\hat{\tau}_1)=1-24\sum_n \frac{n \hat{q}^n}{1-\hat{q}^n},\,  \, E_2(\hat{\tau}_k)=1-24\sum_n \frac{n \hat{q}^{n/k}}{1-\hat{q}^{n/k}},  \\
(E_2(\hat{\tau}_2)/4-E_2(\hat{\tau}_1))\hat{\tau}_1=(-3/4+24\sum_{n,l\geq 1} n \hat q^{n l/2} (\hat q^{n l/2}-1/4))\hat{\tau}_1,
\end{eqnarray}
\begin{eqnarray}
\frac{\pi i}{12}(E_2(\hat{\tau}_k)/k^2-E_2(\hat{\tau}_1))\hat{\tau}_1&=&\frac{\pi i}{12}\left(\frac{1}{k^2}-1+24\sum_{n,l\geq 1}n\left(\hat q^{nl}-\frac{\hat q^{nl/k}}{k^2}\right)\right) \hat{\tau}_1.\nn
\end{eqnarray}


The derivatives of the $E_2$ Eisenstein series can be written as
\begin{eqnarray*}
\frac{d E_2}{d \tau}&=&-48 \pi i\sum_{k,l\geq 1} k^2 l q^{kl},\\
\frac{d^2 E_2}{d \tau^2}&=&96\pi^2\sum_{k,l\geq 1}k^3 l^2 q^{kl},\\
A(q,j)&=&E_2^{(j)}=-24(2\pi i)^j\sum_{k,l\geq 1}k^{j+1}l^jq^{kl}.
\end{eqnarray*}

The function $A(q,j)$ has limit 0 when $q\rightarrow 0$, which is relevant for the behaviour at large cutoff. Using the previous formulae we calculate the third
and fourth contributions to $S_+$ as
\begin{align}
\frac{\partial^3 \tr \rho^n}{\partial  n^3}|_{n=3}=&\frac{(\eta_1\bar{\eta_1})^n}{(\eta_n\bar{\eta_n})}\left(
\left(-S_{UV}+\frac{\pi i}{12}(E_2(\hat\tau_n)/n^2-E_2(\hat \tau_1))\hat \tau_1+c.c.\right)^3\right.\label{tr3}\\
&\left. 3\left(-S_{UV}+\frac{\pi i}{12}(E_2(\hat\tau_n)/n^2-E_2(\hat \tau_1))\hat\tau_1+c.c.\right)  \right.\nn \\
&\times \left.   \left(-\frac{\pi i}{12} E_2'\hat \tau_1^2/n^2-\frac{\pi i}{6}E_2\hat \tau_1/n^3+c.c.\right)          \right. \nn\\
&\left.+\frac{\pi i}{12}E_2''\frac{\hat\tau_1^3}{n^2}+\frac{\pi i}{2}E_2'\frac{\hat\tau_1^2}{n^5}+\frac{\pi i}{2}E_2\frac{\hat\tau_1}{n^4}+c.c. \right)|_{n=3} \nn\\
=&\hat q^{2/9}\prod_n (1-\hat q^n)^4(1+\hat q^{n/3}+\hat q^{2n/3})^2\times \nn\\
&\left((-S_{UV}+\frac{\pi i}{12}(E_2(\hat\tau_3)/3^2-E_2(\hat\tau_1))\hat\tau_1+c.c.)^3\right. \nn\\
& +\left. 3\left(-S_{UV}+\frac{\pi i}{12}(E_2(\hat\tau_3)/3^2-E_2(\hat\tau_1))\hat\tau_1+c.c.\right)\left(-\frac{\pi i}{12} E_2'\hat \tau_1^2/3^2-\frac{\pi i}{6}E_2\hat \tau_1/3^3+c.c.\right) \right.\nn\\
&\left. +\frac{\pi i}{12}E_2''(\frac{\hat\tau_1}{3^2})^3+\frac{\pi i}{2}E_2'\frac{\hat\tau_1^2}{3^5}+\frac{\pi i}{2}E_2\frac{\hat\tau_1}{3^4}+c.c. \right).\nn
\end{align}

\begin{align}
\frac{\partial^4 \tr \rho^n}{\partial n^4}|_{n=4}=&\tr \rho^n\left(\left(-S_{UV}+\frac{\pi i}{12}(E_2(\hat\tau_4)/4^2-E_2(\hat\tau_1))\hat\tau_1\right)^4\right. \label{tr4}\\
&\left. + 4 \left(-S_{UV}+\frac{\pi i}{12}(E_2(\hat\tau_4)/4^2-E_2(\hat\tau_1))\hat\tau_1\right)\times \right.\nn\\
&\left.\left( \frac{\pi i}{12}E_2''\left(\frac{\hat\tau_1}{4^2}\right)^3+\frac{\pi i}{2}E_2'\frac{\hat\tau_1^2}{4^5}+\frac{\pi i}{2}E_2\frac{\hat\tau_1}{4^4}+c.c.\right) \right.\nn\\
&\left.+3\left(-\frac{\pi i}{12}E_2'\left(\frac{\hat\tau_1}{4}^2\right)^2-\frac{\pi i}{6}E_2\frac{\hat\tau_1}{4^3}+c.c.\right)^2\right.\nn\\
+&\left.6\left(-S_{UV}+\frac{\pi i}{12}(E_2(\hat\tau_4)/4^2-E_2(\hat\tau_1))\hat\tau_1+c.c.\right)^2 \left(-\frac{\pi i}{12}E_2'\left(\frac{\hat\tau_1}{4^2}\right)^2
-\frac{\pi i}{6}E_2\frac{\hat\tau_1}{4^3}+c.c.\right)\right.\nn\\
&\left.-\frac{\pi i}{12} E_2'''\left(\frac{\hat\tau_1}{4^2}\right)^4-\pi i E_2''\frac{\hat\tau_1^3}{4^7}-3\pi iE_2'\frac{\hat\tau_1^2}{4^6}-2\pi iE_2\frac{\hat\tau_1}{4^5}+c.c.\right)\nn\\
\tr \rho^4=&\hat q^{5/16}\prod_n(1-\hat q^n)^6(1+\hat q^{n/2})^2(1+\hat q^{n/4})^2.\nn
\end{align}


\bibliographystyle{unsrt}
\bibliography{biblio4}

\def\cprime{$'$} \def\cprime{$'$} \def\cprime{$'$}
\begin{thebibliography}{10}

\bibitem{Obregon:2010zz}
O.~Obreg\'on.
\newblock {Superstatistics and Gravitation}.
\newblock {\em Entropy}, 12:2067--2076, 2010.

\bibitem{PhysRevE.88.062146}
O.~Obreg\'on and A.~Gil-Villegas.
\newblock Generalized information entropies depending only on the probability
  distribution.
\newblock {\em Phys. Rev. E}, 88:062146, Dec 2013.

\bibitem{super}
C.~Beck and E.G.D. Cohen.
\newblock Superstatistics.
\newblock {\em Physica A: Statistical Mechanics and its Applications}, 2003.

\bibitem{obregontorres}
O.~Obreg\'on and J.~Torres-Arenas.
\newblock { H-theorem and Thermodynamics for generalized entropies that depend
  only on the probability. To be published.}

\bibitem{tsallis2}
C.~Tsallis and A.M.C. Souza.
\newblock {Constructing a statistical mechanics for Beck-Cohen
  superstatistics}.
\newblock {\em Physical Review E}, 2004(67):026106, 2003.

\bibitem{Shannon}
C.~E. Shannon.
\newblock {A Mathematical Theory of Communication}.
\newblock {\em The Bell System Technical Journal}, 27:623, 1948.

\bibitem{holzhey}
F.~Holzhey, F.~Larsen, and F.~Wilczek.
\newblock {Geometric and Renormalized Entropy in Conformal Field Theor}.
\newblock {\em Nuclear Physics B}, (424):443--467, 1994.

\bibitem{calabresecardy}
C.~Pasquale and J.~Cardy.
\newblock Entanglement entropy and quantum field theory.
\newblock {\em Journal of Statistical Mechanics: Theory and Experiment},
  2004(06):P06002, 2004.

\bibitem{Calabrese:2009qy}
P.~Calabrese and J.~Cardy.
\newblock {Entanglement entropy and conformal field theory}.
\newblock {\em J.Phys.}, A42:504005, 2009.

\bibitem{Ryu:2006bv}
S.~Ryu and T.~Takayanagi.
\newblock {Holographic derivation of entanglement entropy from AdS/CFT}.
\newblock {\em Phys.Rev.Lett.}, 96:181602, 2006.

\bibitem{Ryu:2006ef}
S.~Ryu and T.~Takayanagi.
\newblock {Aspects of Holographic Entanglement Entropy}.
\newblock {\em JHEP}, 0608:045, 2006.

\bibitem{Hubeny:2007xt}
V.E. Hubeny, M.~Rangamani, and T.~Takayanagi.
\newblock {A Covariant holographic entanglement entropy proposal}.
\newblock {\em JHEP}, 0707:062, 2007.

\bibitem{Nishioka:2009un}
T.~Nishioka, S.~Ryu, and T.~Takayanagi.
\newblock {Holographic Entanglement Entropy: An Overview}.
\newblock {\em J.Phys.}, A42:504008, 2009.

\bibitem{Casini:2011kv}
H.~Casini, M.~Huerta, and R.C. Myers.
\newblock {Towards a derivation of holographic entanglement entropy}.
\newblock {\em JHEP}, 1105:036, 2011.

\bibitem{Fursaev:2006ih}
D.V. Fursaev.
\newblock {Proof of the holographic formula for entanglement entropy}.
\newblock {\em JHEP}, 0609:018, 2006.

\bibitem{Headrick:2010zt}
M.~Headrick.
\newblock {Entanglement Renyi entropies in holographic theories}.
\newblock {\em Phys.Rev.}, D82:126010, 2010.

\bibitem{Lewkowycz:2013nqa}
A.~Lewkowycz and J.~Maldacena.
\newblock {Generalized gravitational entropy}.
\newblock {\em JHEP}, 1308:090, 2013.

\bibitem{Faulkner:2013yia}
T.~Faulkner.
\newblock {The Entanglement Renyi Entropies of Disjoint Intervals in AdS/CFT}.
\newblock 2013.

\bibitem{Hartman:2013mia}
T.~Hartman.
\newblock {Entanglement Entropy at Large Central Charge}.
\newblock 2013.

\bibitem{Maldacena:2013xja}
J.~Maldacena and L.~Susskind.
\newblock {Cool horizons for entangled black holes}.
\newblock {\em Fortsch.Phys.}, 61:781--811, 2013.

\bibitem{Camps:2013zua}
J.~Camps.
\newblock {Generalized entropy and higher derivative Gravity}.
\newblock {\em JHEP}, 1403:070, 2014.

\bibitem{lenzimalacarne}
E.~K. Lenzi, L.~C. Malacarne, and R.~S. Mendes.
\newblock {Path Integral Approach to the Nonextensive Canonical Density}.
\newblock {\em Physica A}, 278:201--213, 2000.

\end{thebibliography}

\end{document}